\documentclass[aip,graphicx]{revtex4-1}

\usepackage{caption}
\usepackage{subcaption}
\usepackage{graphicx}

\draft 

\begin{document}

\preprint{AIP/RSI21-AR-01798}

\title{Beam Particle Identification and Tagging of Incompletely Stripped Heavy Beams with HEIST} 


\author{A.K. Anthony}
\email{anthonya@frib.msu.edu.}
\affiliation{National Superconducting Cyclotron Laboratory, Michigan State University. East Lansing, MI, 48824, USA}
\affiliation{Department of Physics and Astronomy, Michigan State University. East Lansing, MI, 48824, USA}

\author{C.Y. Niu}
\author{R.S. Wang}
\affiliation{National Superconducting Cyclotron Laboratory, Michigan State University. East Lansing, MI, 48824, USA}

\author{J. Wieske}
\affiliation{National Superconducting Cyclotron Laboratory, Michigan State University. East Lansing, MI, 48824, USA}
\affiliation{Department of Physics and Astronomy, Michigan State University. East Lansing, MI, 48824, USA}

\author{K.W. Brown}
\affiliation{National Superconducting Cyclotron Laboratory, Michigan State University. East Lansing, MI, 48824, USA}
\affiliation{Department of Chemistry, Michigan State University. East Lansing, MI, 48824, USA}

\author{Z. Chajecki}
\affiliation{Department of Physics, Western Michigan University. Kalamazoo, MI, 49008, USA}

\author{W.G. Lynch}
\affiliation{National Superconducting Cyclotron Laboratory, Michigan State University. East Lansing, MI, 48824, USA}
\affiliation{Department of Physics and Astronomy, Michigan State University. East Lansing, MI, 48824, USA}

\author{Y. Ayyad}
\altaffiliation{IGFAE, Universidade de Santiago de Compostela, E-15782, Santiago de Compostela, Spain.}
\affiliation{National Superconducting Cyclotron Laboratory, Michigan State University. East Lansing, MI, 48824, USA}

\author{J. Barney}
\affiliation{National Superconducting Cyclotron Laboratory, Michigan State University. East Lansing, MI, 48824, USA}
\affiliation{Department of Physics and Astronomy, Michigan State University. East Lansing, MI, 48824, USA}

\author{T. Baumann}
\affiliation{National Superconducting Cyclotron Laboratory, Michigan State University. East Lansing, MI, 48824, USA}

\author{D. Bazin}
\author{S. Beceiro-Novo}
\affiliation{National Superconducting Cyclotron Laboratory, Michigan State University. East Lansing, MI, 48824, USA}
\affiliation{Department of Physics and Astronomy, Michigan State University. East Lansing, MI, 48824, USA}

\author{J. Boza}
\affiliation{Department of Physics, Western Michigan University. Kalamazoo, MI, 49008, USA}

\author{J. Chen}
\affiliation{National Superconducting Cyclotron Laboratory, Michigan State University. East Lansing, MI, 48824, USA}

\author{K.J. Cook}
\affiliation{National Superconducting Cyclotron Laboratory, Michigan State University. East Lansing, MI, 48824, USA}
\affiliation{Department of Physics and Astronomy, Michigan State University. East Lansing, MI, 48824, USA}

\author{M. Cortesi}
\author{T. Ginter}
\affiliation{National Superconducting Cyclotron Laboratory, Michigan State University. East Lansing, MI, 48824, USA}

\author{W. Mittig}
\author{A. Pype}
\affiliation{National Superconducting Cyclotron Laboratory, Michigan State University. East Lansing, MI, 48824, USA}
\affiliation{Department of Physics and Astronomy, Michigan State University. East Lansing, MI, 48824, USA}

\author{M.K. Smith}
\affiliation{National Superconducting Cyclotron Laboratory, Michigan State University. East Lansing, MI, 48824, USA}

\author{C. Soto}
\affiliation{Department of Physics, Western Michigan University. Kalamazoo, MI, 49008, USA}

\author{C. Sumithrarachchi}
\affiliation{National Superconducting Cyclotron Laboratory, Michigan State University. East Lansing, MI, 48824, USA}

\author{J. Swaim}
\affiliation{Department of Physics, Western Michigan University. Kalamazoo, MI, 49008, USA}

\author{S. Sweany}
\author{F.C.E. Teh}
\author{C.Y. Tsang}
\author{M.B. Tsang}
\author{N. Watwood}
\affiliation{National Superconducting Cyclotron Laboratory, Michigan State University. East Lansing, MI, 48824, USA}
\affiliation{Department of Physics and Astronomy, Michigan State University. East Lansing, MI, 48824, USA}

\author{A.H. Wuosmaa}
\affiliation{Department of Physics, University of Connecticut. Storrs, CT, 06269, USA}



\date{\today}

\begin{abstract}
A challenge preventing successful inverse kinematics measurements with heavy nuclei that are not fully stripped is identifying and tagging the beam particles. For this purpose, the HEavy ISotope Tagger (HEIST) has been developed. HEIST utilizes two micro-channel plate timing detectors to measure time of flight, a multi-sampling ion chamber to measure energy loss, and a high purity Ge detector to identify isomer decays and calibrate the isotope identification system. HEIST has successfully identified $^{198}$Pb and other nearby nuclei at energies of about 75 MeV/A. In the experiment discussed, a typical cut containing 89\% of all $^{198}$Pb$^{+80}$ in the beam had a purity of 86\%. We examine the issues of charge state contamination. The observed charge state populations of these ions are presented and are moderately well described by the charge state model GLOBAL.
\end{abstract}


\maketitle 

\section{Introduction}
\label{into}
At accelerator facilities worldwide, projectile fragmentation is an important mechanism for producing rare-isotope beams (RIBs)  \cite{gade2016nscl, anne1987achromatic,geissel1992gsi,kubo1992riken}. Projectile fragmentation produces a wide variety of isotopes, from stable to short-lived, that can be subsequently filtered by fragment separators to select the desired species. Unlike stable beam facilities, RIBs produced at fragmentation facilities typically have lower intensities, larger emittances, and more than one isotopic species, even after passing through a fragment separator.

For many experiments, it is necessary to identify the beam on an event-by-event basis.  Often, an appropriate choice of separator settings can remove most unwanted ions in the beam. Subsequently measuring the time of flight (ToF) from the fragment separator to the experimental setup allows the remaining rare isotopes in the beam to be uniquely identified (i.e. tagged). Such identification becomes more difficult when these secondary beams are heavy and the beam particles have overlapping velocity distributions. In this case, a ToF measurement is not sufficient to resolve the beam species. Beam species can still be distinguished by an additional measurement of the energy deposited in an ion chamber, $\Delta$E.

Heavy beams are routinely tagged by the $\Delta$E-ToF-B$\rho$ method at the RIBF \cite{kubo1992riken} and GSI \cite{geissel1992gsi} facilities. At those facilities, beams energies are $E/A \ge 200$ MeV, and beam ions are often fully stripped. Slower beams, where the ions are not fully stripped complicate the picture. At the energy regime of interest, $E/A\approx$ 74 MeV, such heavy ions typically retain their 1s electrons, making them ``helium-like" nuclei with  Q/e = Z-2, but with non-negligible  probabilities for them to be ``lithium-like" ions with Q/e = Z-3 or ``hydrogen-like" with Q/e = Z-1. The HEavy ISotope Tagger (HEIST) was developed at the National Superconducting Cyclotron Laboratory (NSCL) to tag and track secondary projectile fragments at energies where ions are not fully stripped.

This leads to three questions: (1) how accurately can we tag these ions? (2) how serious are the complications from multiple charge states in these ions? and (3) how accurately do theoretical calculations predict the performance of our beam identification setup and the production of different charge states? Answers to these questions can facilitate future efforts to tag heavy nuclei.

\section{HEIST Setup and Operation}
\label{setup}

\begin{figure*}[htbp!]
\centering
		\includegraphics[width=0.9\textwidth]{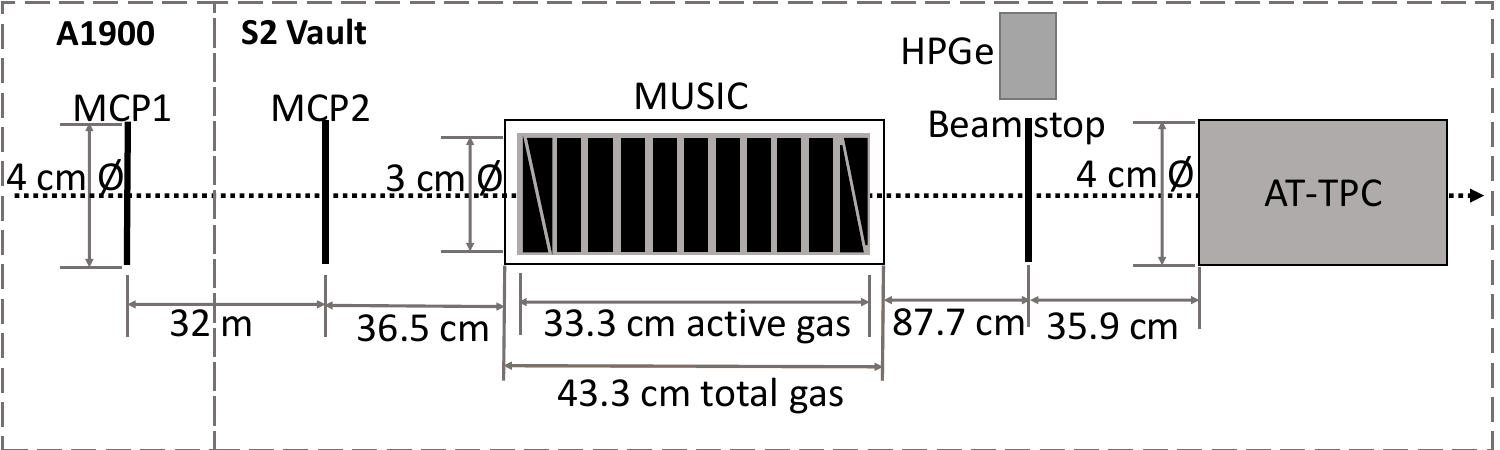}
		\caption{Detector locations along the beamline. Not to scale.}
		\label{fig:beamline}
\end{figure*}

HEIST was coupled to the A1900 fragment separator \cite{morrissey2003commissioning} at the NSCL and consists of three distinct detector systems. The detector systems are: a time of flight (ToF) system composed of two micro-channel plates (MCPs), a Multi-Sampling Ionization Chamber (MUSIC) that provides beam tracking and energy loss ($\Delta$E) information, and a high-purity Ge (HPGe) crystal used to detect isomer $\gamma$-rays. The physical locations of the detectors and the experimental end station, here the Active-Target Time Projection Chamber (AT-TPC) \cite{bradt2017commissioning}, are shown schematically in Fig. \ref{fig:beamline}. The location of the first MCP (MCP1) at the A1900 focal plane is shown in Fig. \ref{fig:A1900}. Further details of each of the detector systems can be found in Sec. \ref{detectors}. 

\subsection{A1900 Setup}
\label{A1900Setup}

\begin{figure*}[htbp!]
\centering
		\includegraphics[width=0.9\textwidth]{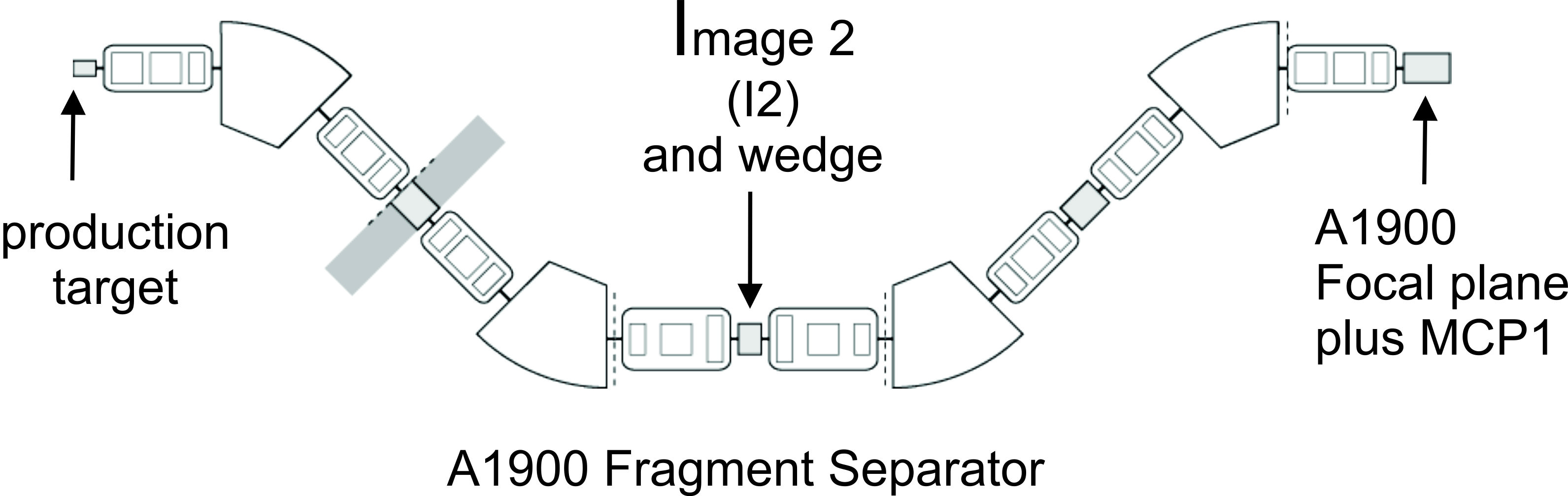}
		\caption{ A1900 Fragment Separator: The locations of the production target, Image 2 dispersive focus and the 1900 focal plane are shown.}
		\label{fig:A1900}
\end{figure*}

At the Coupled Cyclotron Facility (CCF) of the NSCL \cite{gade2016nscl,morrissey2003commissioning}, a primary $^{208}$Pb beam was accelerated first in the K500 Cyclotron, striped and injected into the K1200 Cyclotron where it was accelerated to its final energy of E/A = 85 MeV. Rare isotopes were then produced by bombarding the $^{208}$Pb beam on a $^{9}$Be production target, with an areal density of 22.5 mg/cm$^2$, located outside of the K1200 Cyclotron \cite{morrissey2003commissioning}. This produces a mixture of primary and secondary beams that enter the A1900 fragment separator \cite{morrissey2003commissioning}, shown in Fig.  \ref{fig:A1900}. 

Midway through the A1900 magnetic separator, there is a dispersive focus at Image2 (I2) where the beam particles are focused to different points depending on their rigidity. Using a set of thin narrow slits at I2 to define a 2 mm aperture, we selected ions that have a magnetic rigidity of $B\rho$ = 3.271 Tm with a precision of $\pm~0.035\%$. This rigidity transmits $^{192}$Hg ions with Q/e = 77 at an energy of E/A = 79.7 MeV, as well as other ions with a similar rigidity.

The isotopic selection was improved by passing the remaining ions through a 23.5 mg/cm$^2$ Polyimide foil to degrade the energy. Ions with different atomic numbers emerge from the foil with different rigidities \cite{morrissey2003commissioning}; slits at the second dispersive focus at the A1900 focal plane removed unwanted ions with magnetic rigidities that differ from the desired rare isotope \cite{morrissey2003commissioning}. This second set of slits was set to select particles with rigidity of $B\rho$ = 3.117 Tm $\pm~0.18\%$ corresponding to $^{192}$Hg ions with a charge of Q/e = 78 by blocking undesired isotopes with larger or smaller magnetic rigidities.

The selected isotopes then pass through the first MCP (MCP1) that served as a start detector for the ToF system and go on to the second MCP (MCP2), placed 32 m downstream of MCP1 that served as the stop detector, as shown in Fig. \ref{fig:beamline}. Using the two MCPs we measured the ToF and the velocity of the beam particles before they passed into the MUSIC detector and the AT-TPC.

\subsection{$\Delta$E-ToF-B$\rho$ Method}
\label{dEToFBrho}
The momentum cut at the dispersive focus of the A1900 separator is related to the A1900 rigidity setting, the beam momentum, beam charge and mass as follows:

\begin{equation}
    B\rho = \frac{p}{q} \simeq \frac{mv}{q},
\end{equation} 

 Here $p$ is momentum, $q$ is charge, $m$ is mass, and $v$ is velocity. The energy loss of a charged particle through a material, $\Delta$E, is related to its velocity and atomic number by 
 
 \begin{equation}
     \Delta E \approx C\frac{Z^2}{v^2}f(v)~,
 \end{equation}
 where $f(v)$ is a slowly varying function of the velocity $v$ \cite{Bethe1953}.
 
 Combining the energy loss in the ion chamber, $\Delta$E, with the velocity determined by ToF provides uniquely the atomic number, Z, via Eq. 2. Combining the velocity with the magnetic rigidity via Eq. 1 provides the ratio of mass to charge. For reactions where only one charge state is present for each isotope, this allows a unique determination of the mass and atomic numbers of each particle. For slower heavy fragments where multiple charges are probable, one can take guidance from calculations of the charge state distributions and beam optics calculations that indicate where each charge state should appear in the particle identification (PID) spectrum. In Sec. $\ref{qContamination}$, we test these charge state models via measurements of characteristic isomer fed $\gamma$-rays that uniquely identify the beam fragment independent of charge states. 

\subsection{A1900 Simulation}
\label{simulation}

 To design HEIST, we modeled its performance with the LISE++ beam simulation program \cite{tarasov2008lise++}. The width of the I$_2$ slit was optimized by considering the trade-off between momentum resolution and beam rate. The rms time resolution $\delta t$ of an MCP can be estimated by the  relationship $\delta t = \tau_{rise} V_{noise}/V_{signal}$ where  $\tau_{rise}$ is the rise time of the MCP anode signal, $V_{noise}$ is the rms noise and $V_{signal}$ is the amplitude of the MCP anode signal. From this we estimated that we could achieve a FWHM time of flight resolution of about 300 to 400 ps. Allowing for different path lengths for different ions, we conservatively estimated a time resolution of 500 ps could be achieved. LISE predicted the energy loss straggling in the MUSIC with 300 torr CF$_4$ to be somewhat less than 0.7\% (FMHW). Using the software package SRIM/TRIM \cite{ziegler2010srim}, we calculated that an energy loss resolution of 0.85 \% (FWHM) could be achieved.
 
 \begin{figure}[htbp!]
\centering
	\includegraphics[width=0.8\linewidth]{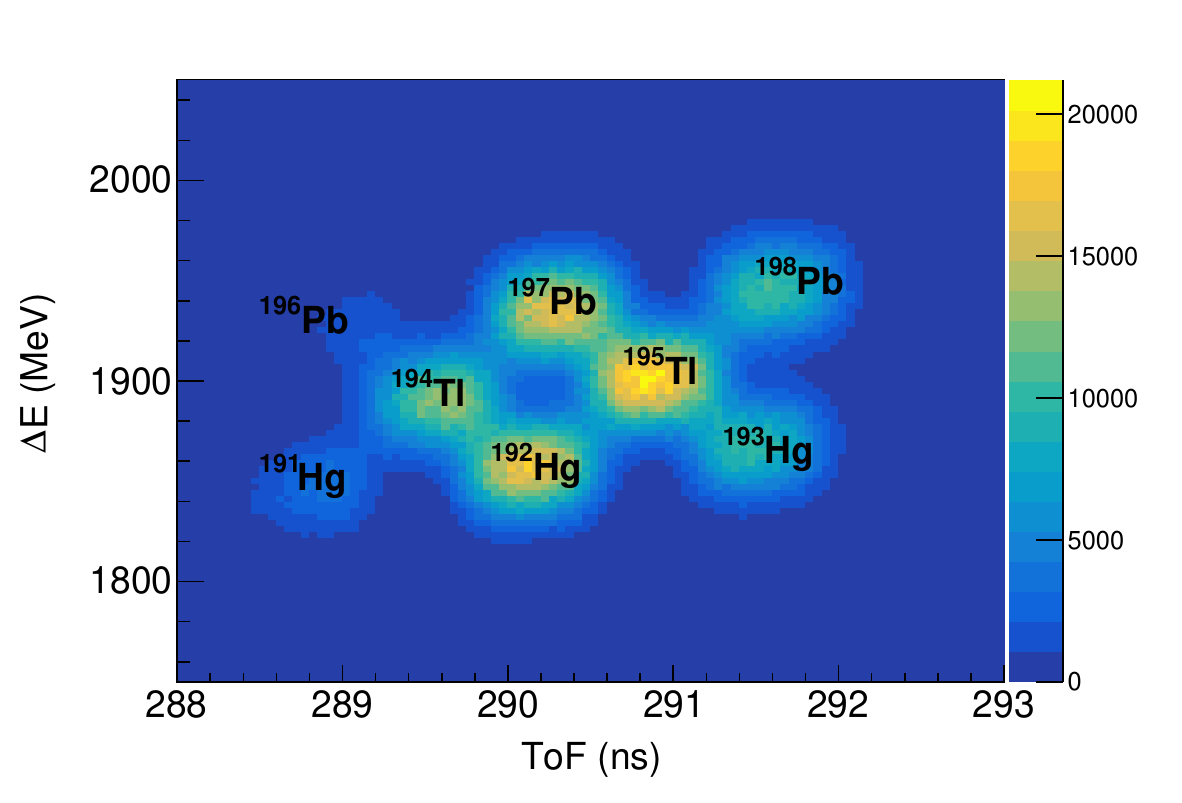}
	\caption{The LISE++ simulated PID of several secondary beam isotopes. The charge states are helium-like.}
	\label{fig:pid}
\end{figure}

 The beam optics and detector response were simulated with LISE++ to determine our predicted ability to successfully identify beam particles. Fig. \ref{fig:pid} shows the $\Delta$E vs ToF PID plot simulated with LISE++ assuming the A1900 configuration discussed in Sec. \ref{A1900Setup}. For the simulation, we assumed a conservative $\Delta$E resolutions of 2 \% (FWHM), significantly larger than the pure energy straggling calculations made with SRIM or LISE++, to include fluctuations from secondary effects, such as high energy delta electrons that escape the detector volume.  With these assumptions, we predicted that the various isotopes in the beam could be adequately resolved.

\section{HEIST Construction and Operation}
\label{detectors}

\subsection{MCP}
\label{mcpDesign}

The timing detector was based on a thin emitting foil, 0.5 $\mu$m polypropylene, placed along the path of the beam with an MCP readout. These MCPs provide excellent timing resolution and can operate at a rate exceeding 1 MHz without damage. Using thin foils avoids unnecessary degradation of the beam emmittance.

\begin{figure}[htbp!]
\centering
		\includegraphics[width=0.8\linewidth]{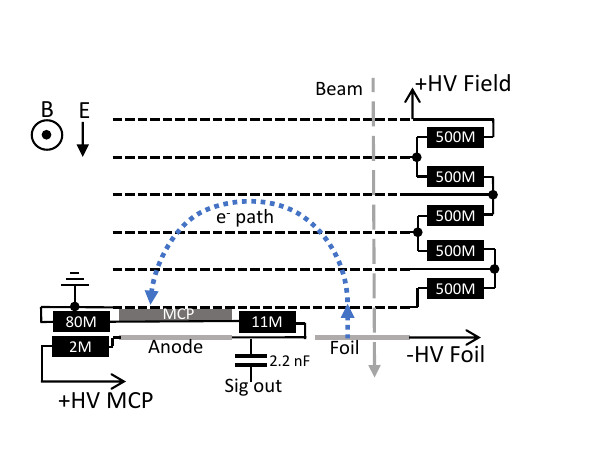}
		\caption{(Color online) Schematic of the working principle of the MCP detector, not to scale}
		\label{fig:MCPschematic}
\end{figure}

The design of both MCP timing detectors is similar to that described in Refs. \cite{wiggins2017development, wiza1979microchannel}. Fig.  \ref{fig:MCPschematic} illustrates the design and operating principles of this MCP timing detector. When a beam particle passes through the thin foil, electrons are released by ionizing atoms which are accelerated away from the foil by a static electric field. The trajectories of the electrons are bent by electric and magnetic fields through an angle of $180^\circ$ so that they hit a stack of 40 mm diameter MCP electron amplifiers produced by Photonis. The two MCPs in the stack are arranged front to back in a chevron configuration to multiply the primary ionization electrons by a gain factor of approximately $10^6$. The amplified charges are collected onto an anode plate located behind the MCP. The multiplication process in the MCP is fast, and the corresponding signals have a rise time of about 3 ns.

A series of electrodes, shown by the dashed black lines in Fig. \ref{fig:MCPschematic}, are connected with a resistor chain. When biased, they generate the electric field that, in conjuncture with a permanent magnetic field, bends the electrons. The highest potential electrode was biased to $V_{E-field}\approx2$ kV. The permanent magnetic field is generated by attaching high-field, permanent neodymium magnets on two nickel-plated iron plates that sandwich the MCP detectors. The locations of the permanent magnets were positioned so that the magnetic field is  approximately 100 G and uniform to within 2 \% over the electron trajectories. In addition, a -1 kV bias is applied to the foil and a positive bias of $V_{mcp} \approx 2.5$ kV is applied to the MCPs and anode.

Signals from the anode were amplified using a non-inverting 200 gain fast amplifier (Ortec FTA820A). The signals were then discriminated using a constant fraction discriminator (Canberra 454) and passed to a TDC (Caen 1190). The signals from MCP1 at the A1900 focal plane were discriminated near the detector, and then sent to the S2 vault at the NSCL where MCP2, the MUSIC, and the TDC to digitize the signals were located.

\subsection{MUSIC}
\label{musicDesign}

\subsubsection{Mechanical Design}
\begin{figure}
    \centering
    \begin{subfigure}[t]{.8\linewidth}
    \centering
    \includegraphics[width=.8\linewidth]{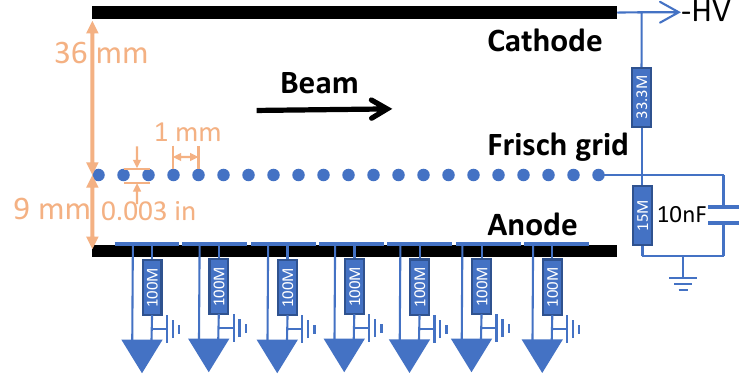}
    \caption{MUSIC schematic} \label{fig:IcSchematic}
    
    \end{subfigure}
    \begin{subfigure}[b]{.8\linewidth}
    \centering
    \includegraphics[width=.8\linewidth]{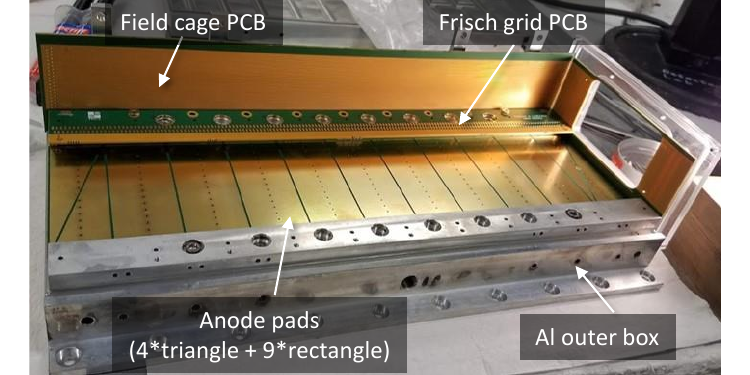}
    \caption{MUSIC picture} \label{fig:IcPicture}
    \end{subfigure}
    
    \caption{(Color online) The MUSIC used in this experiment. (\subref{fig:IcSchematic}) shows a schematic of MUSIC detector, not to scale. (\subref{fig:IcPicture}) is a photo of the MUSIC during construction. The first two and last two segments are triangular pads used for measuring beam position within the chamber via charge division.}
\end{figure}

The MUSIC detector was constructed to measure the energy loss of the secondary beam particles. Fig. \ref{fig:beamline} shows an overview indicating the horizontal dimensions of the MUSIC chamber and the physical layout of the anode pads of the MUSIC chamber. Fig. \ref{fig:IcSchematic} shows a schematic side view of the MUSIC chamber indicating the orientation of the anode electrodes, Frisch grid, and cathode.  This MUSIC detector is loosely based on the MUSIC80 design employed at GSI to measure the energy loss of heavy beam particles \cite{schneider2000technical}. 

Fig. \ref{fig:IcPicture} is a photograph of the inside of the MUSIC taken during construction and the anode structure is visible. The MUSIC detector has an active area of 62 mm $\times$ 36 mm and active length of 333 mm. The anode and cathode are both printed circuit boards (PCBs) (333 mm $\times$ 128 mm), mounted onto aluminum plates that are held parallel to each other and separated by 45 mm. The PCBs in this detector were fabricated with a 1.6 mm thick halogen free substrate, IT-170RGA1 and their 1oz copper electrode surfaces were coated with Electroless Nickel Immersion Gold  (ENIG) plating. The cathode consists of a single conducting (332 mm $\times$ 127 mm) pad.

The anode is split into 9 rectangular segments, with two additional anode segments (the first and the last) that are subdivided into pairs of triangle pads resulting in a total of 13 pads, as shown in Fig. \ref{fig:beamline} and \ref{fig:IcPicture}. The rectangular middle nine anode segments are 30 mm $\times$ 82 mm conducting surfaces separated from each other by 1 mm insulating gaps. The triangular pads (24 mm $\times$ 82 mm) provide information about the horizontal beam positions near the entrance and exit of the chamber using the technique of charge division. The beam tracking capabilities of the detector are discussed in detail in Sec. \ref{anode pads}.

Each of the thirteen anode pads were individually amplified via Zepto Systems CSA-60-100 charge sensitive preamplifiers and a programmable spectroscopy amplifier (CAEN n568). The slow output from the spectroscopy amplifier is digitized by a peak sensing ADC (Mesytec MACD-32). The fast output from the spectroscopy amplifier was discriminated using a leading edge discriminator (Philips 7106) and digitized using a TDC (CAEN V1190). The typical signal to noise ratio for the anode signal was 0.07 \%.

The Frisch grid was fabricated using 76 $\mu$m diameter beryllium copper wires and oriented parallel to, and 9 mm above, the anode. The pitch between neighboring wires was chosen to be 1 mm and is discussed in detail in Sec. \ref{frishGrid}. 

The electric field inside the drift region of the MUSIC detector is generated by the electrode surfaces of the PCBs and of the exit and entrance windows. The side wall PCBs have 0.2 mm wide conducting strips spaced 1mm apart and connected through a resistor chain to the cathode and anode providing a uniform electric field near the side walls that extends from the cathode to the Frisch grid and between the Frisch grid and the anode. In addition, the beam enters and exits the MUSIC field cage through 0.5 $\mu$m thick polypropylene windows with an active areas of 30 mm $\times$ 30 mm. By evaporating 1.6 mm wide Al strips, spaced 2.54 mm apart, and biasing them to the same voltage as the side wall PCB we ensured a uniform electric field throughout the MUSIC field cage. The locations of the first and last rectangular anode sections were chosen so that all electrons that drift to these pads pass through a region where the drift field is uniform to better than 1\%. 

The drift field is defined by the negative high voltage $V_c\ge-5~kV$ on the cathode and the ground potential of the anode pads which are individually grounded through 100 M$\Omega$ resistors. A simplified voltage divider circuit and Frisch grid filter is shown in Fig. \ref{fig:IcSchematic}. An iseg VHQ 203M power supply was used to bias the cathode. Table \ref{tab:gasProperties} contains the bias voltage, drift field, and electron drift velocity for gas configurations used in this paper which are discussed in detail in Sec. \ref{gasSelection}.

\begin{table*}[ht!]
  \begin{center}
    \caption{Selected gas properties of CF$_4$ at running conditions.}
    \label{tab:gasProperties}
    \begin{tabular}{|c|c|c|c|c|} 
    \hline
       & Cathode  &  &  $e^-$ drift  & Avg energy per \\
      Gas (torr)& voltage (V) & $E_C$ (V/cm) & velocity (cm/$\mu$s) & ion pair (eV) \\
        \hline
            CF$_4$ 150 & -1000  & 192   & 10.3  & 34.3 \\
        \hline
            CF$_4$ 294.5 & -1000  & 192  & 8.3  & 34.3 \\
        \hline
    \end{tabular}
  \end{center}
\end{table*}

This light-weight MUSIC assembly sits within a larger aluminum box that provides additional shielding and serves as a pressure vessel. To minimize gas contamination and recombination in the active region of the detector, gas flows directly from the gas handling system into the active region through numerous small holes in the anode PCB, seen in Fig. \ref{fig:IcPicture}. This enforces a laminar gas flow from the anode to the cathode and minimizes recombination. Then the gas flows through small holes in the cathode to the outer pressure holding vessel and finally back to the gas handling system.

\subsubsection{Frisch Grid Considerations}
\label{frishGrid}

Beam particles entering the gas volume of the MUSIC detector make a straight line path ionizing the gas and creating a cylindrical distribution of electrons and positive ions along the path of the beam particle. Since the electron drift velocity greatly exceeds that of the ions, positive ions remain in the gas long after the electrons are collected on the anode. Absent a Frisch grid, this positive ion charge induces a negative image charge on the anode that reduces the net anode signal by an amount that depends on the distance of the track from the anode. With a Frisch grid in place, the image charges mainly appear on the Frisch grid instead, and the anode signal is basically independent of the vertical position of the track within the chamber.

Following Bunemann \cite{bunemann1949design}, we define the grid ``inefficiency", $\gamma$, to be a measure of how independent the field strength at the anode, $E_A$,  is the field created by positive ions in the cathode region, $E_Q$:

\begin{equation}
    \gamma = \frac{dE_A}{dE_Q},
\end{equation}

where $\gamma$ is given by Bunemann's expression: 

\begin{equation}
    \gamma = \frac{d}{2\pi (x_A + r - \pi r^2/d)} \ln \frac{d}{2\pi r}.
\end{equation}

Here, $x_A=9$ mm is the distance between the Frisch grid and the anode PCB boards, $d=1$ mm is the wire pitch, and $r=38~\mu$m is the radius of the Frisch grid wires, shown in Fig \ref{fig:IcSchematic}. This provides a value of $\gamma$ = 0.026, effectively suppressing the dependence of the anode signals on the position of the positive ions in the detector.

If the anode were fully shielded, then no electrons would make it past the grid to the anode. The shielding of the anode to the positive space charge must be balanced against the need for most electrons to reach the anode. The proportion of electrons collected on the anode is roughly equal to the proportion of field lines that bypass the Fricsh grid, and terminate on the anode. Bunemann calculated the condition for the relative field strengths in the anode, $E_A$, and cathode, $E_C$, drift regions where all field lines by-pass the Frish grid \cite{bunemann1949design}:

\begin{equation}\label{eq:fieldCon}
    \frac{E_A}{E_C} \ge \frac{1+2\pi r/d}{1-2\pi r/d}.
\end{equation}

For our chosen values of $d$ and $r$, the field in the anode region must be ~1.6 times larger than the field in the cathode region to satisfy this condition. We constructed a voltage divider for the field cage accordingly to achieve $E_A/E_C~=~1.8$, satisfying the condition given by Eq. \ref{eq:fieldCon}, and minimizing the loss of signal due to electrons terminating on the Frisch grid instead of the anode. 

\subsubsection{Position sensitivity of the MUSIC}
\label{anode pads}

The four triangular anode pads  allow the horizontal ``x"-position of the track to be determined at the entrance and exit of the MUSIC detector. Each track deposits charge on the pad surfaces that are directly above it. This makes the deposited charge in the triangular pad proportional to the width of the pad at the x-position where the charge is deposited. Defining beam right to be positive x direction and the left pad to be the one that is wider on the beam-left side of the anode plane, one can  divide the difference in the charges in the right and left pads by the sum of both charges to get the x-position as follows:

\begin{equation}
    x = x_{max} \frac{Q_R - Q_L}{Q_R + Q_L},
\end{equation}

\noindent where $Q_L$ and $Q_R$ is the charge deposited on the left and right triangular pads, respectively, and $x_{max}$ = 41 mm is half the width of the anode plane.

The y-position of the track can also be determined from the time difference between the signal in MCP2 located 40 cm upstream, which records the arrival time of the beam particle and the anode signal which starts when the ionized electrons pass through the Frisch grid. If the drift velocity of electrons in the detection gas is known, then the delay time for the anode signal is a linear function of the drift distance from the track to the anode plane. By combining the x and y positions determined by these anode signals,the location and direction of beam tracks can be reconstructed in 3D space and extrapolated to different points along the beamline.

\subsubsection{Gas selection and properties}
\label{gasSelection}
CF$_4$ was selected as the detection gas. It has the virtues that it is not explosive, has a fast drift velocity, which minimizes pile-up effects, and a large stopping power. Table \ref{tab:gasProperties} provides relevant counter gas properties for CF$_4$, including the electron drift field strengths, drift velocities calculated using MagBoltz \cite{BIAGI1999234}, and average energy required to produce an ion/electron pair \cite{reinkingCF4}.

\subsection{HPGe}
\label{otherDesign}

 For calibration and characterization of the HEIST system, individual isotopes in the beam were uniquely identified by stopping them in an aluminum block after they pass through the MUSIC detector and detecting $\gamma$-rays fed by the decay of long-lived isomers in these isotopes. These $\gamma$-rays were detected in an OR120 Ortec High Purity Germanium crystal (HPGe), which is a liquid-nitrogen cooled GEM series HPGe coaxial detector system. The Ge crystal in this model of detector has dimensions 80.8 mm $\times$ 105.8 mm, with a 4 mm gap between the end cap of the housing and the crystal. A 1 mm thick Aluminum layer and 0.7 mm of inactive Germanium served as absorbing layers. Despite the attenuation of some  $\gamma$-rays in the absorbing layers, we were able to match the observed coincident $\gamma$-rays to known isomers fed $\gamma$-rays and thereby uniquely identify many of the isotopes in the beam, as discussed in Sec. \ref{beamID}. At typical gamma-ray energies observed in this experiment (300~-~1000 keV), a resolution of 1.1 keV (FWHM) was achieved.

\section{Commissioning HEIST}
\label{pb}

A number of stable primary $^{208}$Pb beams were used to commission HEIST, ensuring proper functionality and calibration of the various detector systems described above. The properties of these beams are detailed in Table \ref{tab:stableBeam}. 

\begin{table}[ht!]
  \begin{center}
    \caption{Selected properties of stable $^{208}$Pb beams used to test detectors.}
    \label{tab:stableBeam}
    \begin{tabular}{|c|c|c|c|c|} 
    \hline
      Beam type & $B \rho$ (Tm) & Charge states \\
        \hline
            Pilot beam  & 4.4785  & 63$^+$\\
        \hline
            Calibration beam  & 2.5782  & 79$^+$, 80$^+$, 81$^+$, 82$^+$\\
        \hline
    \end{tabular}
  \end{center}
\end{table}

\subsection{ToF Calibration}
\label{tofCal}

\begin{figure}
\centering
\begin{subfigure}[t]{0.8\linewidth}
\centering
\includegraphics[width=\textwidth]{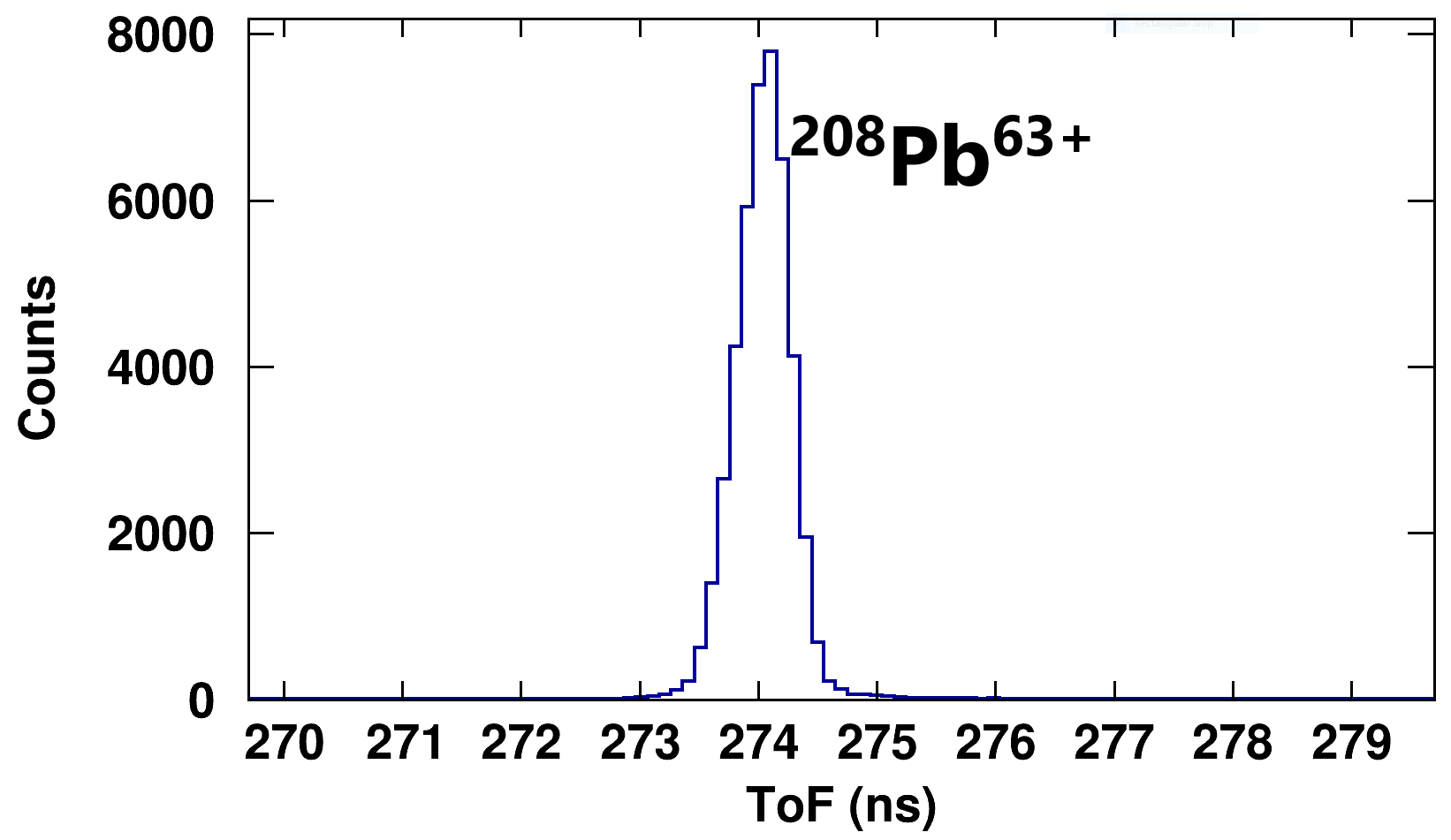} 
\caption{Undegraded $^{208}$Pb$^{63+}$ beam.} \label{fig:primary}
\end{subfigure}

\begin{subfigure}[t]{0.8\linewidth}
\centering
\includegraphics[width=\textwidth]{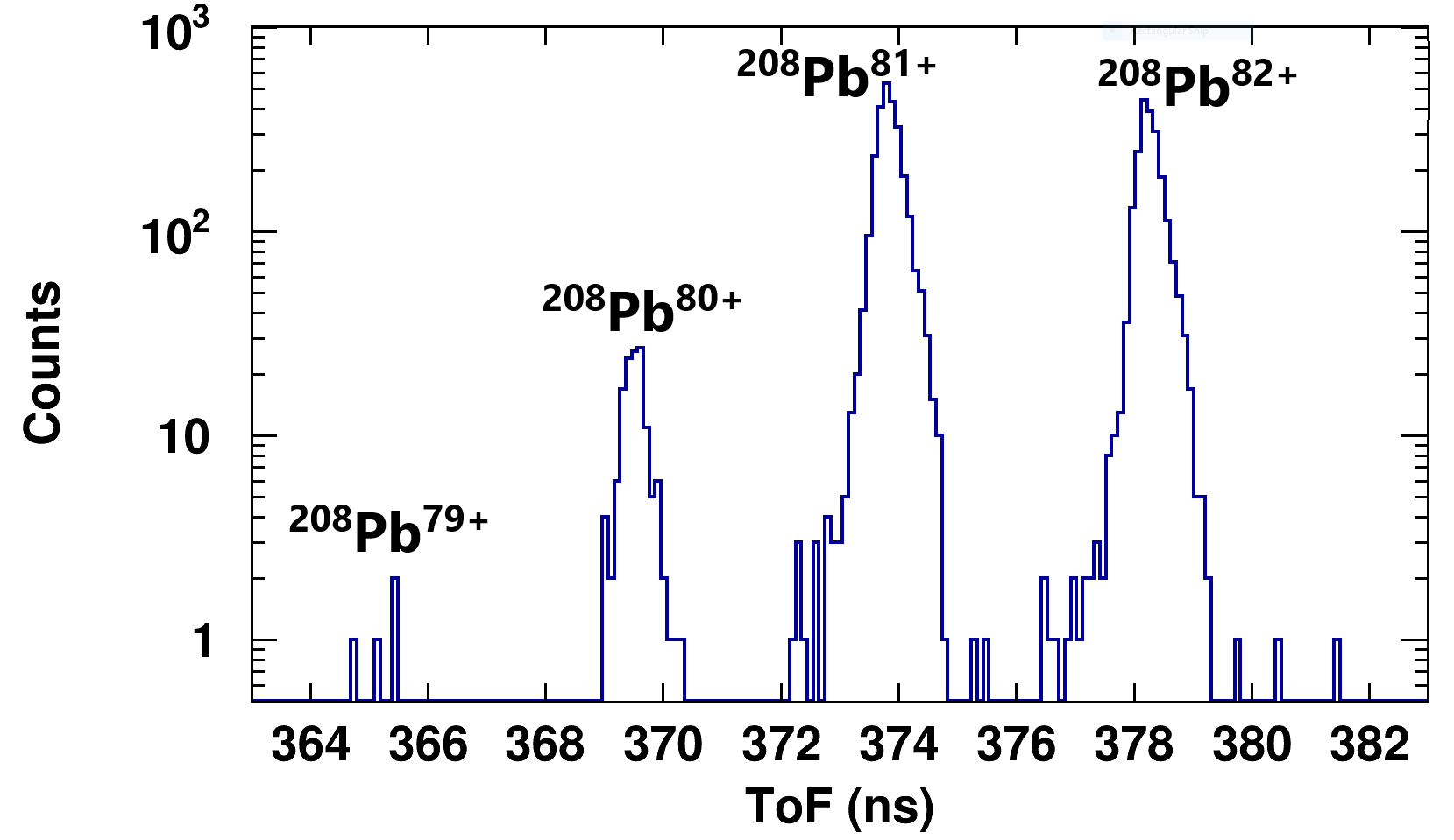} 
\caption{Multiple charge states of $^{208}$Pb.} \label{fig:calibration}
 \end{subfigure}

 \caption{ToF spectra for two primary beam settings. Shown in (\subref{fig:primary}) is the ToF spectra for an undegraded primary $^{208}$Pb$^{63+}$ beam with B$\rho=4.4785$ Tm. Shown in (\subref{fig:calibration}) is the ToF spectra for multiple charge states of the $^{208}$Pb beam with B$\rho=2.5782$ Tm.}

\end{figure}

The absolute time of flight between the two MCP detectors was determined using a $^{208}$Pb$^{+63}$ primary beam from the CCF. The resulting ToF spectrum for this beam is shown in Fig. \ref{fig:primary}. The beam's rigidity, and therefore velocity, is set by the A1900. Using LISE++, the physical ToF for the primary beam can be calculated for the flight path along the beamline. By fitting the ToF spectra, we corrected for unknown cable delays so that the absolute flight times were obtained for all beams. 

To provide a further check, the primary beam was degraded prior to the A1900 and transmitted with the magnetic rigidity set to 2.5782 Tm. As the primary beam passes through the degrader, it will exchange electrons with the target producing multiple charge states which after the B$\rho$ selection of the A1900 will have different velocities. We observe the degraded primary beam in four charge states: Q/e=Z, Z-1, Z-2 and Z-3, where Z=82. The ToF values shown in Fig. \ref{fig:calibration} for these four charged states agree with the expected values to within 0.2 ns.

\section{ Rare Isotope Beam Measurements}
\label{rib}
 
\subsection{Beam Identification}
\label{beamID}

\begin{figure}[htbp!]
	\includegraphics[width=0.9\linewidth]{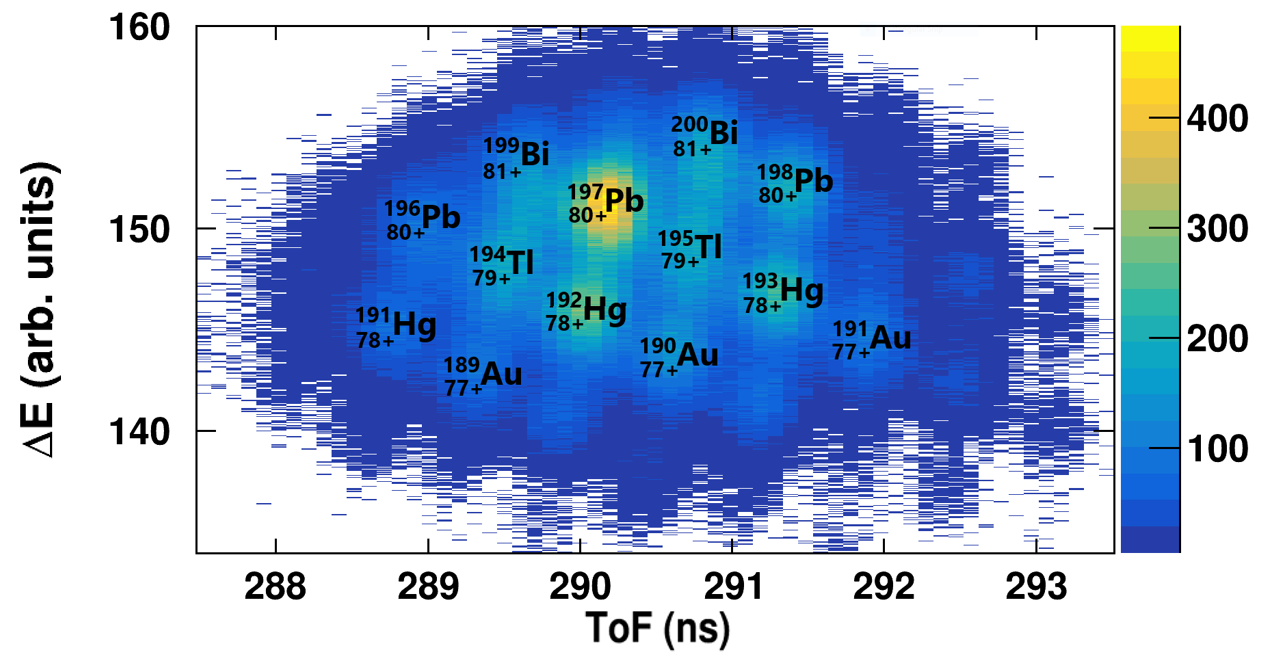}
	\caption{The PID plot of secondary beam isotopes.}
	\label{fig:pidl}
\end{figure}

Fig. \ref{fig:pidl} shows the correlation between the secondary beam ToF (x-axis) and the sum of the energy losses measured by the nine rectangular anode electrodes in the MUSIC for the secondary beam tuned  to $B\rho$ = 3.1173 Tm. Multiple peaks appear in this two dimensional spectrum. Each represents a different isotope, which is marked on the figure. Since this PID plot is unchanged by stopping the beam ions in an aluminum block approximately 30 cm downstream, we can identify these ions by measuring the $\gamma$-rays emitted after an isomeric state decay with the HPGe detector.

\begin{figure}
\centering
\begin{subfigure}[t]{0.8\linewidth}
\centering
\includegraphics[width=\textwidth]{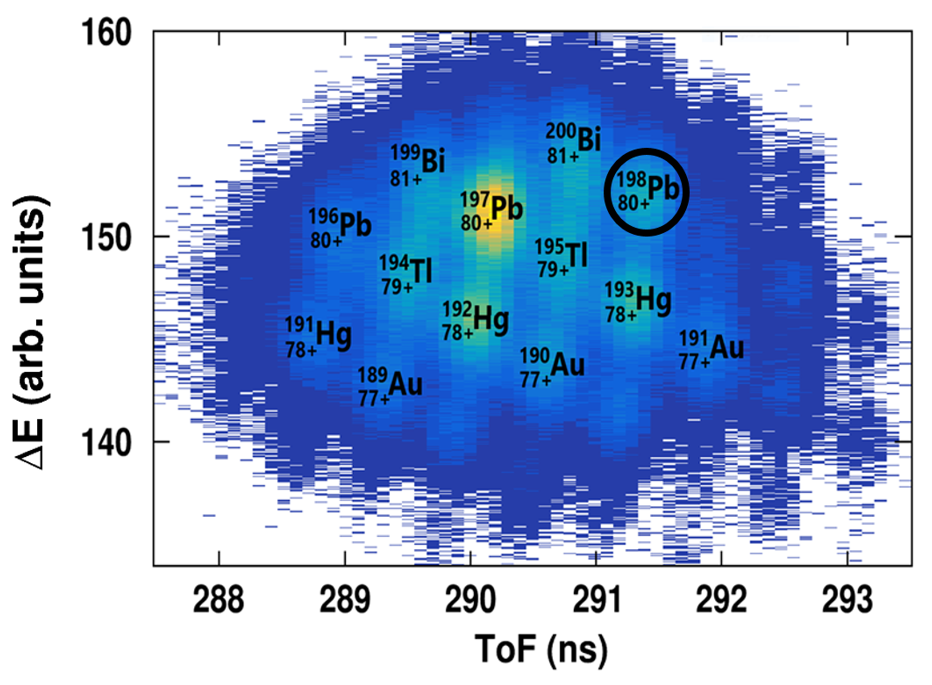} 
\caption{PID with gate on $^{198}$Pb$^{80+}$.} \label{fig:PID_gated}
\end{subfigure}

\begin{subfigure}[t]{0.8\linewidth}
\centering
\includegraphics[width=\textwidth]{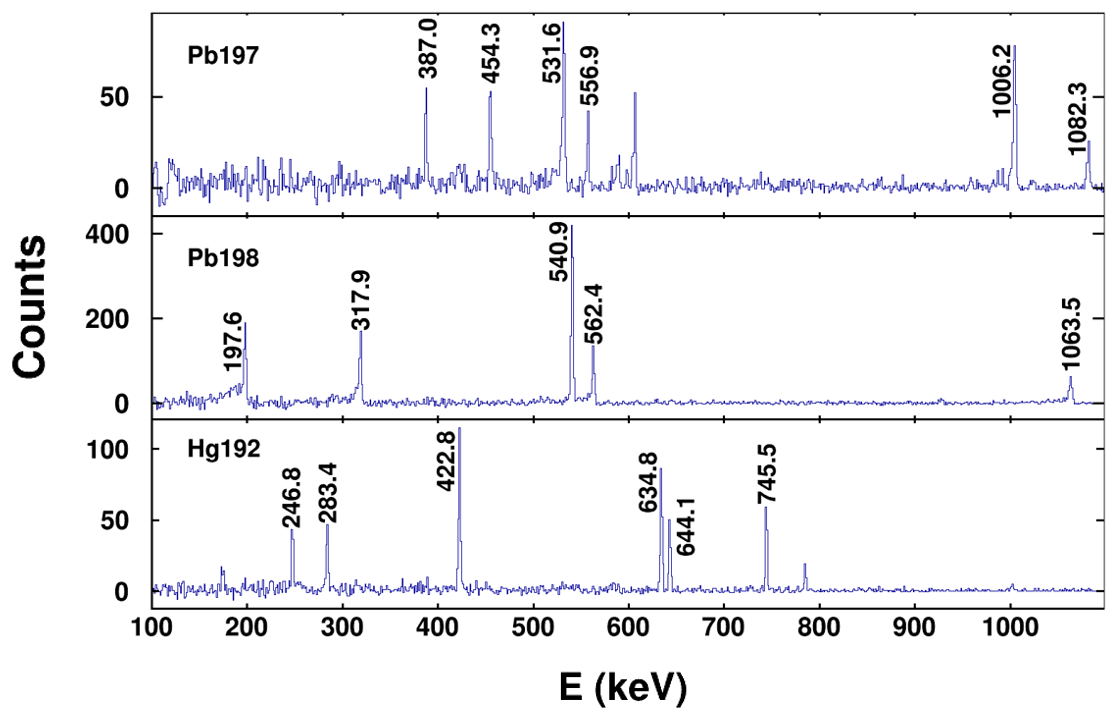} 
\caption{Gamma spectra for various nuclei.} \label{fig:gamma}
 \end{subfigure}

 \caption{Shown in (\subref{fig:PID_gated}) is the secondary beam PID spectra illustrating a gate on $^{198}$Pb. Shown in (\subref{fig:calibration}) are gamma spectra gated on the PID spectra for the specific nuclei: $^{197}$Pb (top), $^{198}$Pb( middle), and $^{192}$Hg (bottom).}
\label{fig:isotopeID}
\end{figure}

 Fig. \ref{fig:isotopeID} illustrates the identification of secondary beam isotopes. Gates on each of the PID peaks and coincident $\gamma$-ray spectra were obtained for each peak. The black contour in Fig. \ref{fig:PID_gated} surrounds the $^{198}$Pb PID peak with Q/e=80. The middle panel of Fig. \ref{fig:gamma} shows the $^{198}$Pb $\gamma$-ray peaks that were observed for ions detected within the black contour. For isomer decays to feed these $\gamma$-ray  transitions, $^{198}$Pb isomer states must have long enough lifetimes so that some isomers survive the 590 ns flight from the production target. 

Fig. \ref{fig:gamma} also shows some other examples of gamma ray spectra that enabled us to identify other isotopes. Table  \ref{tab:tagged_isotopes} lists all the nuclei that were identified via isomer fed delayed $\gamma$-rays \cite{sonzogni2005nudat} and their associated energies. While our single HPGe detector does not provide details of the feeding of isomer decays into the delayed $\gamma$-ray we observe, there is enough consistent information obtained from all of these decays that there can be no doubt about isotope assignments shown in Fig. \ref{fig:pidl}. 

\begin{table*}[ht!]
  \begin{center}
    \caption{A list of isotopes that have been confirmed by isomer tagging, their corresponding charge states measured in the PID, and some characteristic gamma rays seen in the HPGe.
    }
    \label{tab:tagged_isotopes}
    \begin{tabular}{l|c|c} 
      \textbf{Isomer-tagged Isotopes}  & \textbf{Typical Gammas (keV)} & \textbf{Charge States Identified} \\
      \hline
      $^{196}$Pb & 337.3~,~371.9~,~689.0 & 79+~,~80+ \\
      $^{197}$Pb & 387.0~,~531.6~,~589.0 & 80+ \\
      $^{198}$Pb & 197.6~,~317.9~,~450.9 & 80+~,~81+ \\
      $^{192}$Hg & 283.4~,~422.8~,~745.5 & 78+ \\
      $^{194}$Hg & 280.2~,~427.9~,~636.3 & 78+\\
      $^{190}$Au & 232.3~,~854.9~,~427.8 & 77+ \\
      $^{191}$Au & 264.0~,~420.1~,~725.2 & 77+~,~78+\\
      $^{199}$Bi & 145.7~,~499.6~,~1002.2 & 81+\\
      $^{201}$Bi & 185.8~,~411.7~,~967.4 & 81+\\
    \end{tabular}
  \end{center}
\end{table*}

\subsection{Beam Resolution}
\label{beamres}

\begin{figure*}
\centering
\begin{subfigure}[t]{0.49\textwidth}
\centering
\includegraphics[width=\textwidth]{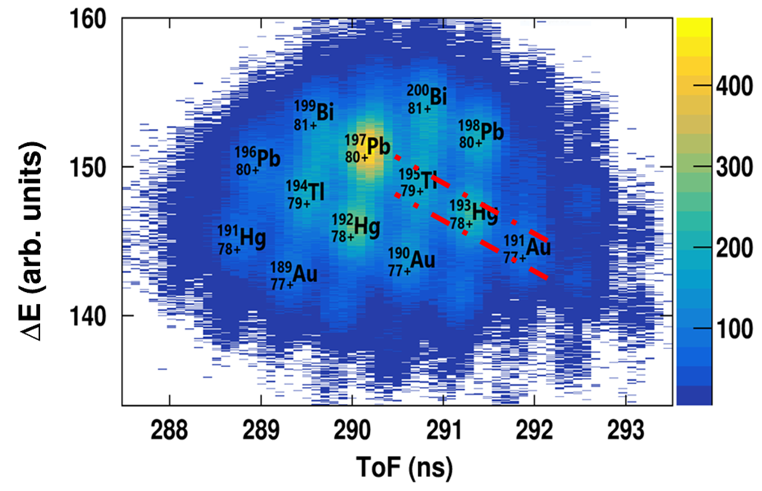} 
\caption{PID with gate for ToF.} \label{fig:ToF_PID_gated}
\end{subfigure}
\begin{subfigure}[t]{0.49\textwidth}
\centering
\includegraphics[width=\textwidth]{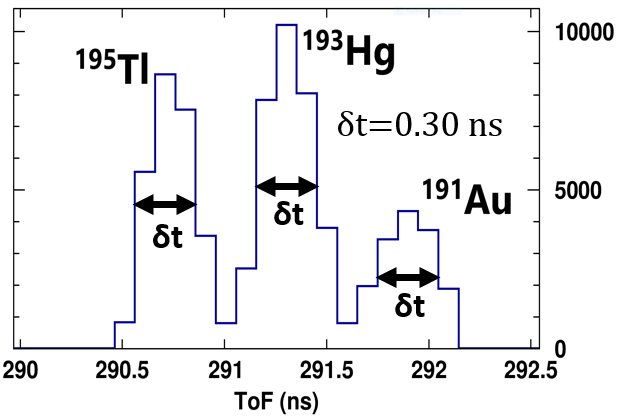} 
\caption{ToF spectrum.} \label{fig:ToF_resolution}
 \end{subfigure}

 \caption{Shown in (\subref{fig:PID_gated}) is the secondary beam PID spectrum with a gate (in red) that provides information about the ToF resolution. Shown in (\subref{fig:calibration}) is the ToF spectrum for the gated data.}

\end{figure*}

Fig. \ref{fig:pid} visually displays the amount of separation between isotopes achieved with HEIST. To make more concrete comparisons to our estimated ToF performance, we first put a gate on the PID spectrum shown by the red dashed line in Fig. \ref{fig:ToF_PID_gated} and plot the enclosed data as a function of the ToF in Fig. \ref{fig:ToF_resolution}. 

The time resolution for each peak is about 300 ps, which indicates that each of these secondary beams follows its own tightly defined trajectory so that the final time resolution is consistent with the simple estimate based on the noise and the rise times for the MCP signals made in Sec. \ref{simulation}.

\begin{figure*}[hbtp!]
\centering
\begin{subfigure}[t]{0.49\textwidth}
\centering
\includegraphics[width=\textwidth]{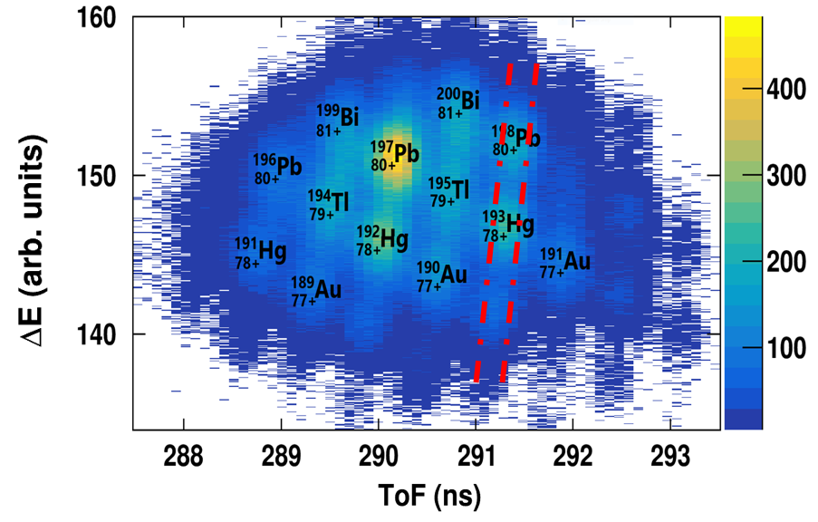} 
\caption{PID with gate for $\Delta$E.} \label{fig:dE_PID_gated}
\end{subfigure}
\begin{subfigure}[t]{0.49\textwidth}
\centering
\includegraphics[width=\textwidth]{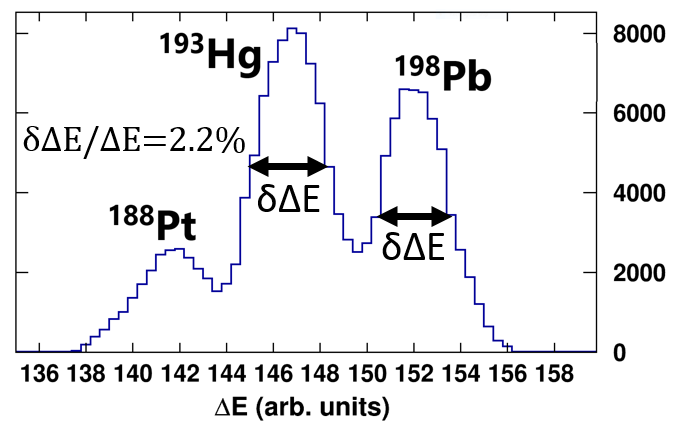} 
\caption{$\Delta$E spectrum.} \label{fig:dE_resolution}
 \end{subfigure}

 \caption{Shown in (\subref{fig:PID_gated}) is the secondary beam PID spectrum with a gate (in red) that provides information about the $\Delta$E resolution. Shown in (\subref{fig:calibration}) is the $\Delta$E spectrum for the gated data.}

\end{figure*}

 To make similarly concrete comparisons to our estimated $\Delta E $ resolution, we apply the nearly vertical gate on the PID spectrum shown in Fig. \ref{fig:dE_PID_gated}, and plot the enclosed data as a function of the the energy loss $\Delta$E in Fig. \ref{fig:dE_resolution}. The observed $\Delta$E resolution is approximately 2.2 \% (FWHM) of the average  $\Delta$E deposited in the MUSIC chamber for that isotope. This is about 2.5 times larger than the estimated $0.85~\%$ resolution given by the calculated straggling using SRIM. To explore this discrepancy, a GEANT4 \cite{AGOSTINELLI2003250, Allison1610988, ALLISON2016186} Monte-Carlo simulation was made using the physical detector geometry to explore secondary effects, like delta electron production. The simulation showed an energy resolution of 3 \% (FWHM), consistent with, though somewhat larger than, the experimentally determined resolution.
 
 It is impossible to quantify the performance of HEIST in a single number. Because the tails of the energy and time distributions of different species overlap, the purity of an isotopic selection depends on the cut made. The maximum purity of a cut is determined by the amount of charge state contamination from same Z isotopes. If there is no directly overlapping charge state in the acceptance of the beamline, a cut can be made that has an arbitrarily high purity, at the cost of lost statistics. The typical $^{198}$Pb$^{80+}$ cut shown in Fig. \ref{fig:ToF_PID_gated} contains 89\% of all $^{198}$Pb$^{80+}$ in the beam, and has a purity of 86\%. A purity of $\ge99\%$ $^{198}$Pb$^{80+}$ was achieved using a cut containing less than 20\% of the all $^{198}$Pb$^{80+}$ in the beam.

\begin{table}[ht!]
  \begin{center}
    \caption{Theoretical ToF calculated by LISE++ compared to the measured ToF after absolute time calibration of the MCPs.}
    
    \label{tab:isotope_tof}
    \begin{tabular}{|c|c|c|} 
    \hline
      \textbf{Isotope} & \textbf{LISE++ ToF ($ns$)} & \textbf{Data ToF (ns)} \\
        \hline
            $^{198}$Pb$^{80+}$ & 292.14 & 292.06 \\
        \hline
            $^{197}$Pb$^{80+}$ & 290.88 & 290.86 \\
        \hline
            $^{196}$Pb$^{80+}$ & 289.61 & 289.68 \\
        \hline
            $^{196}$Pb$^{79+}$ & 292.75 & 292.64 \\
        \hline
            $^{193}$Hg$^{78+}$ & 292.07 & 291.98 \\
        \hline
            $^{192}$Hg$^{78+}$ & 290.77 & 290.74 \\
        \hline
    \end{tabular}
  \end{center}
\end{table}

\subsection{Charge State Contamination}
\label{qContamination}

One of the challenges of data from a RIB containing high-Z isotopes is the issue of charge state contamination, or the presence of more than one charge state configuration for a given isotope in the beam. The acceptance of our beam line limits the number of charge states of a given isotope to one or two. Most isotopes arrive at the MUSIC in a helium-like configuration with Q/e = Z-2. However, that does not prevent contamination from another isotope with a less probable charge state from contributing to the PID spectrum within the same PID gate. 

After identification of the isotopic species through isomer tagging in the HPGe, LISE++ was used to calculate ToF for several different charge states of each isotope. Table \ref{tab:isotope_tof} shows a comparison of the calculated ToF from LISE++ and the corrected experimental data. The measured  and calculated ToF agree to within 1 ns, whereas the ToF difference between adjacent charge states for a given isotope is on the order of several ns, allowing for the unique identification of the charge state. 

\begin{figure}[htbp!]
\centering
\begin{subfigure}[t]{0.8\linewidth}
\centering
\includegraphics[width=\textwidth]{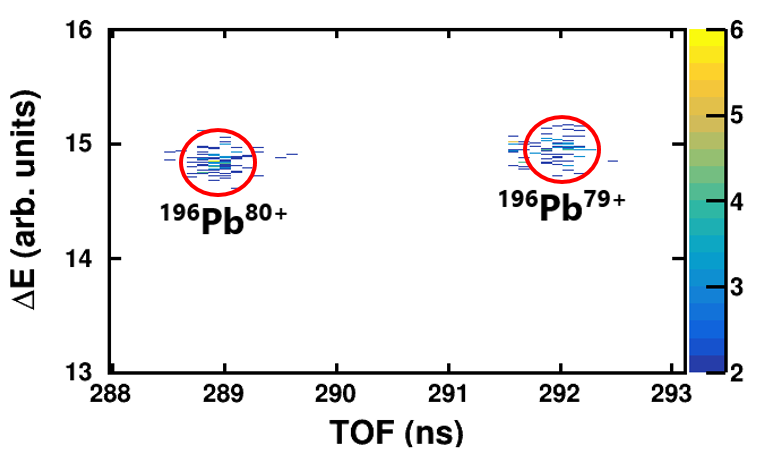} 
\caption{PID with gate on $^{196}$Pb $\gamma$-rays.} \label{fig:PID_qState_gated}
\end{subfigure}

\begin{subfigure}[t]{0.8\linewidth}
\centering
\includegraphics[width=\textwidth]{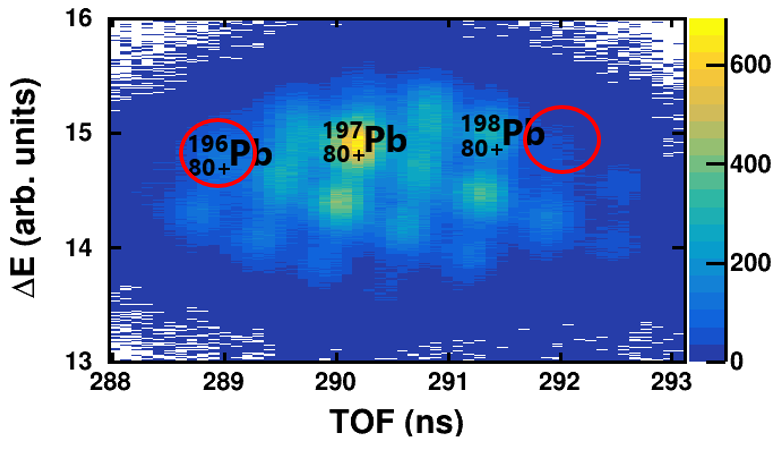} 
\caption{Ungated PID.} \label{fig:PID_qState_full}
\end{subfigure}

 \caption{An example of the presence of multiple charge states in the RIB for $^{196}$Pb. The main 80+ charge state is clearly visible in the ungated PID, while the 79+ state is not obvious. (\subref{fig:PID_qState_gated}) shows the the measured PID for events with a gamma ray fed by the decay of an isomer state in $^{196}$Pb. (\subref{fig:PID_qState_full}) shows the full PID with the location of $^{196}$Pb charge states circled in red.}
\label{fig:charge_state}
\end{figure}

Examples of both lithium-like and hydrogen-like secondary charge states have been detected and identified using the HPGe.  Separation of neighboring charge states in ToF allows for the observed ratio of charge states to be measured accurately. Fig. \ref{fig:charge_state} illustrates this process when the HPGe selection is set to select $^{196}$Pb. The two resulting peaks in Fig. \ref{fig:PID_qState_gated} indicate the presence of two separated charge state peaks, determined to be the helium-like (80+) and lithium-like (79+) configurations by ToF. The peak from the lithium-like charge state is not immediately obvious in the ungated PID, as seen in Fig. \ref{fig:PID_qState_full}. The observed ratio between charge states of the same isotope depends on the acceptance of the beamline and may differ from the actual production ratio at the A1900.

\begin{figure}[htbp!]
    \centering
	\includegraphics[width=0.8\linewidth]{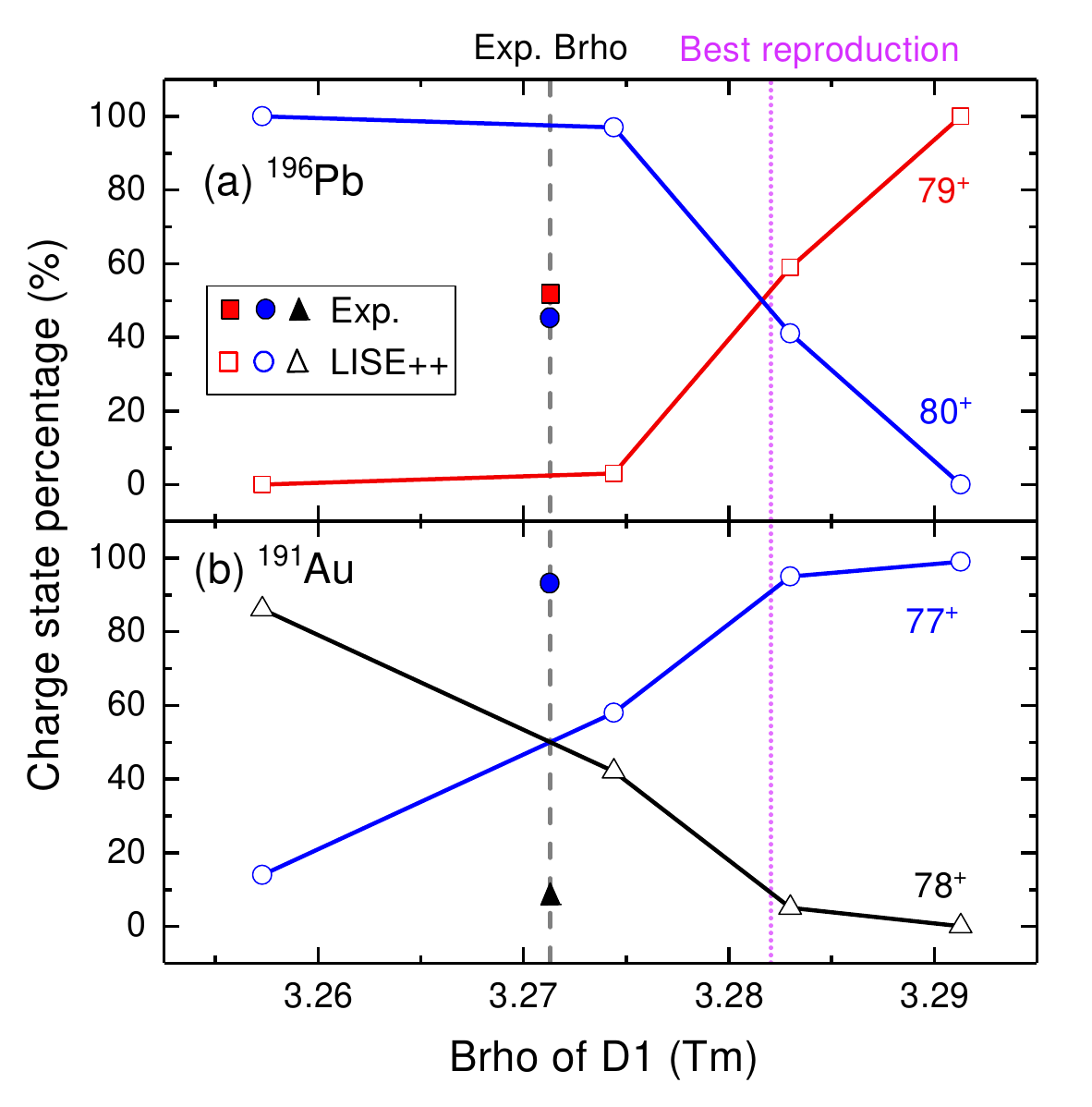}
	\caption{The measured charge state distributions of  (a) $^{196}$Pb, and (b) $^{191}$Au are plotted as a function of rigidity in the A1900. The solid symbols indicate the experimental values, while the open symbols shows the LISE++ simulation results using the GLOBAL model.}
	\label{fig:LISE++_charge_state}
\end{figure}

Fig. \ref{fig:LISE++_charge_state} shows measured charge state distributions for the two most common charge states of $^{196}$Pb and $^{191}$Au. The physical rigidity cut of the A1900 is shown by a vertical bar. Using the GLOBAL \cite{SCHEIDENBERGER1998441} model implemented in LISE++, the theoretical charge state distribution across a range of A1900 rigidity cuts was calculated, as seen in Fig. \ref{fig:LISE++_charge_state}. 

There is a discrepancy between the calculated charge state distributions and experimental measurement. The model reproduces the data better if the B$\rho$ value of the A1900 and beamline is adjusted upwards in the calculation by 0.3\% to 3.283 Tm from 3.271 Tm. This discrepancy is larger than the uncertainty in the calibration of the A1900 rigidity which is believed to be good to $\pm 0.1\%$. This discrepancy can be explained by GLOBAL under-predicting the electron capture cross-sections for these nuclei. Regardless of the reason for the discrepancy, the predictions of GLOBAL are consistent with all measurements made if the adjusted B$\rho$ value of 3.283 Tm is used, as seen in Table \ref{tab:charge_state}.

\begin{table*} [thpb]
  \centering
  \caption{A comparison between the theoretical and measured charge state distributions of isotopes Pb, Hg, and Au with the charge state $Q = Z-3$, $Z-2$, and $Z-1$. The theoretical charge state distributions were calculated using the GLOBAL model with LISE++. The experimental charge state distributions were extracted by gating on the corresponding $\gamma$-rays from known isomers.}
  
  \label{tab:charge_state}
    \begin{tabular}{c|ccc|ccc}
    \hline \hline
      Isotopes  &   & Calculation    & & & Experiment &  \\
                & $Z-3$ & $Z-2$ & $Z-1$ & $Z-3$ & $Z-2$ & $Z-1$ \\
    \hline
      $^{196}$Pb    &59\%   &41\%   &       &52\%~$\pm$~3\%   &48\%~$\pm$~3\%   & \\
      $^{197}$Pb    &       &100\%  &       &       &100\%  & \\
      $^{198}$Pb    &       &100\%  &       &       &99\%~$\pm$~2\%   &1\%~$\pm$~1\% \\
      $^{191}$Hg    &27\%   &73\%   &       &-       &-       &- \\
      $^{192}$Hg    &       &100\%  &       &       &100\%  & \\
      $^{193}$Hg    &       &100\%  &       &-       &-       &- \\
      $^{189}$Au    &2\%    &98\%   &       &-       &-       &- \\
      $^{190}$Au    &       &100\%  &       &       &100\%  & \\
      $^{191}$Au    &       &95\%   &5\%    &       &95\%~$\pm$~5\%   &5\%~$\pm$~1\% \\
    \hline \hline
    \end{tabular}
\end{table*}

\subsection{Beam Tracking}
\label{beamTrack}

The MUSIC detector has the capability to measure the position and direction of the beam as it passes through the detector. As discussed in Sec. \ref{anode pads}, the x-position is determined by charged division with the triangular pads at the entrance and exit of the MUSIC detector and the y position is determined from the drift time of the electrons to the Frisch grid. 

\begin{figure} [htpb] 
  \centering
  \includegraphics [width=0.8\linewidth] {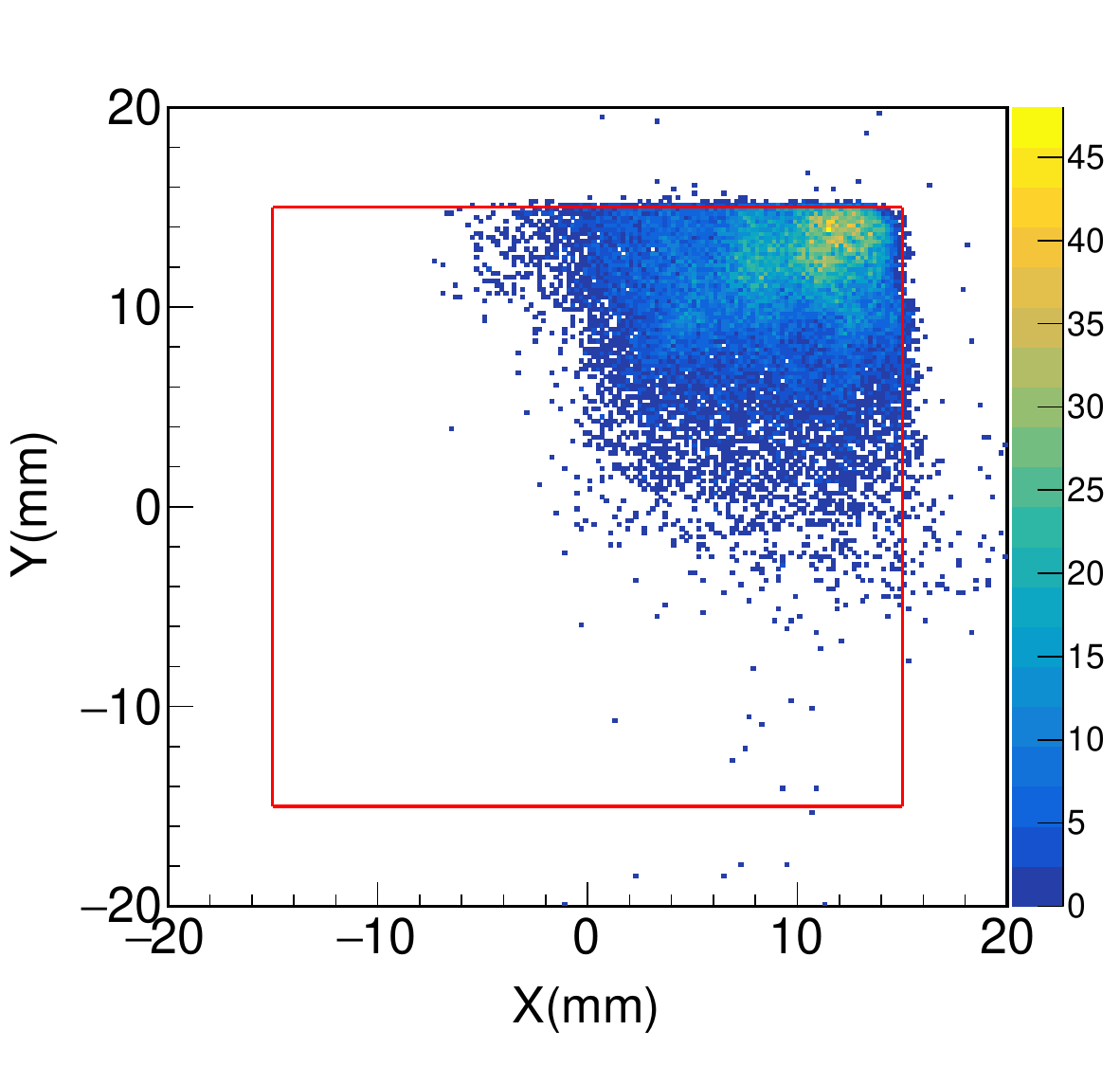}
  \caption{(Color online) Experimental beam hit pattern on the entrance of the MUSIC. The red square shows the edge of the MUSIC window.}
  \label{fig:hit_pattern}
\end{figure}

\begin{figure} [htpb] 
  \centering
  \includegraphics [width=.9\linewidth] {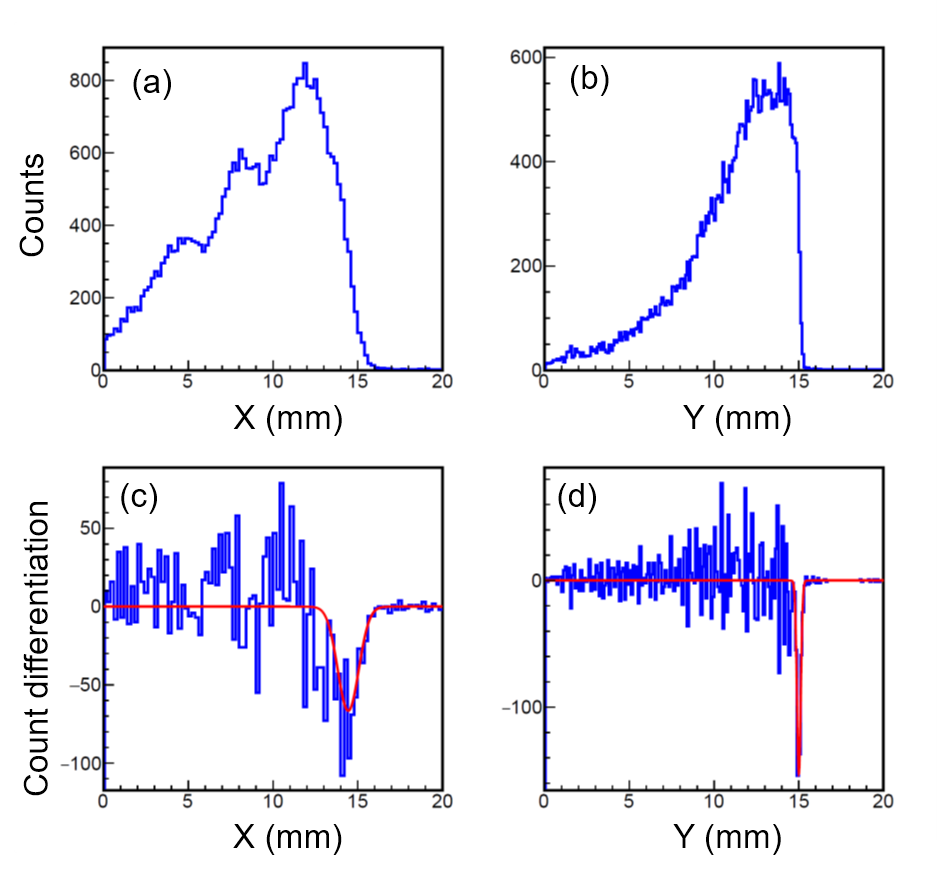}
  \caption{
  (Color online) The position distribution is shown in X direction in (a) and the Y direction in (b). The numerical differentiation of the position spectrum with the edges fitted with a Gaussian function are used to determine the position resolution. (c) shows the differentiated position for X, and (d) shows the differentiated position for Y.}
  \label{fig:position_resolution}
\end{figure}

To check the x- and y- position resolution, the last quadrupole magnet along the beamline was turned off so the beam would clip the edges of the MUSIC entrance window. Fig. \ref{fig:hit_pattern} shows the measured beam profile. The MUSIC entrance window's right and upper edges are clearly shown. The corresponding x- and y- position distribution is shown in the upper panels of Fig. \ref{fig:position_resolution}. A numerical differentiation of the spectrum was performed and fit with a Gaussian in order to determine the position resolution, shown in the lower panels of Fig. \ref{fig:position_resolution}. The obtained x-position resolution was 1.45 mm, while the y-position resolution was 0.27 mm. 

\begin{figure} [htpb] 
  \centering
  \includegraphics [width=0.9\linewidth] {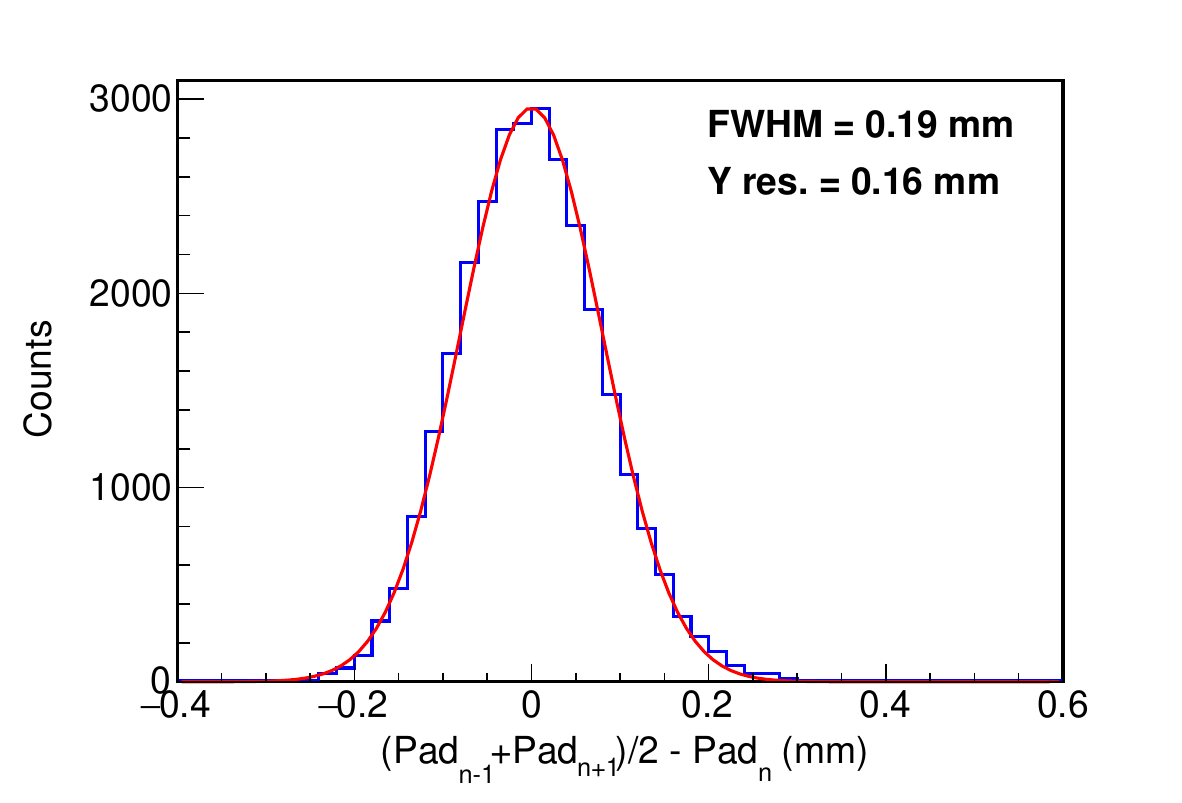}
  \caption{The Y position difference between the average of $n+1$ $\&$ $n-1$ and $n$ rectangle pads. }
  \label{fig:Y_res}
\end{figure}

The estimation of the y position resolution can be improved. There are 9 independent measurements of the y position, one from each rectangular anode pad. A more accurate y position resolution can be obtained by taking the difference between the y positions of neighboring pads. Fig. \ref{fig:Y_res} shows the y position difference between the average of the $(n+1)^{th}$ $\&$ $(n-1)^{th}$ pads, and $n^{th}$ pad. The FWHM of this difference distribution is 0.19 mm, so the y resolution for each pad is $0.19/\sqrt{3/2}=0.16$ mm.

\begin{figure} [htpb] 
  \centering
  \includegraphics [width=1\linewidth] {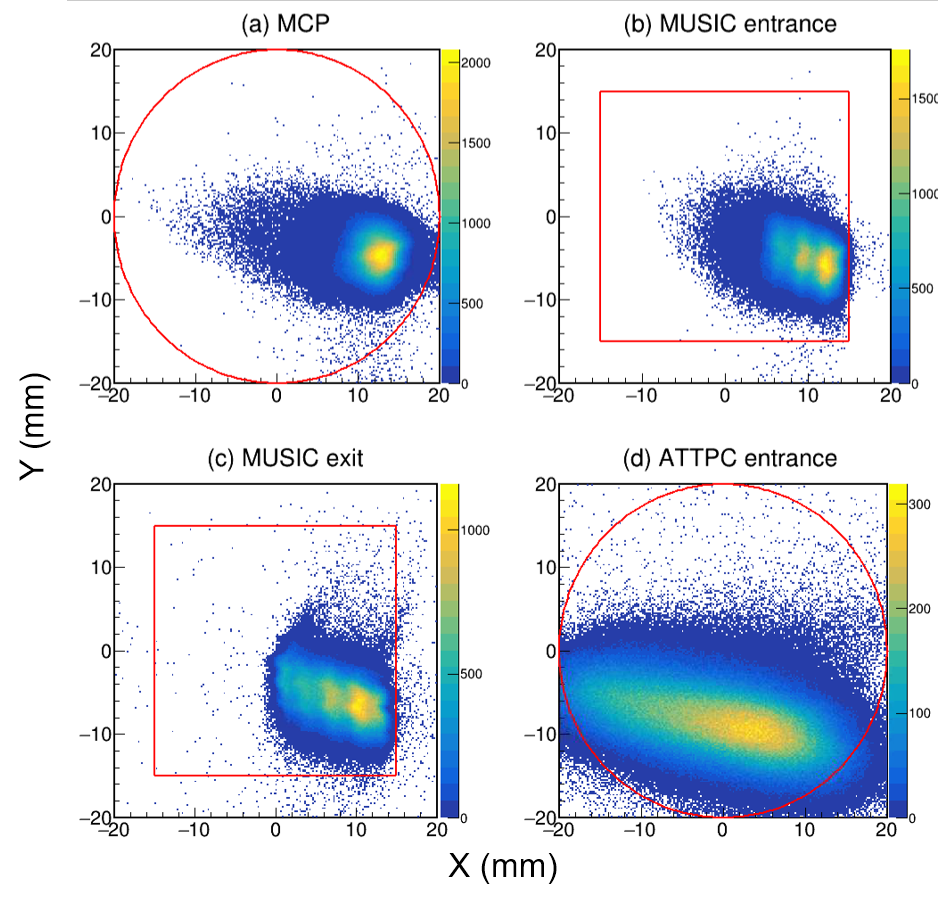}
  \caption{(Color online)  (a) shows the extrapolated position spectrum at MCP2. (b) shows the position spectrum at the entrance of the MUSIC. (c) shows the position spectrum at the exit of the MUSIC. (d) shows the extrapolated position spectrum at the entrance to the AT-TPC. The red lines show the edge of detector windows.}
  \label{fig:beam_tracking}
\end{figure}

As seen in Figs. \ref{fig:beam_tracking} and \ref{fig:position_resolution}, within the MUSIC detector there is a structure to the x-position of the beam. Even though the momentum selection at I2 is very precise, the wedge foil modifies the energy of the beam and therefore its rigidity, and different beam particles ultimately enter the MUSIC at different x positions leading to the structure observed.

With accurate measurements of the positions at the entrance and exit window of the MUSIC, the beam particle trajectory can be  extrapolated to other detectors along the beamline. Fig. \ref{fig:beam_tracking} shows the beam tracking results with hit patterns at different positions. At MCP2, before the MUSIC, the beam was well focused. In the MUSIC the beam was off-center, indicating a slight misalignment of the apparatus. The beam nonetheless enters the acceptance of the ATTPC entrance window 1.2 m downstream of the MUSIC detector. 

\clearpage

\section{Conclusion}
\label{conclusion}
The HEIST system was designed, built, and commissioned. It consists of two MCP detectors, a MUSIC ion chamber, and a HPGe detector. HEIST was able to successfully tag and track secondary isotopes produced from the fragmentation of a 85 MeV/A $^{208}$Pb beam. The time of flight performance of this system betters the design criterion, but the charge resolution, while acceptable, is somewhat worse than the design value. GEANT4 calculations for the charge resolution are consistent with the measured charge resolution, indicating that the larger measured energy loss resolution results from secondary processes such as delta electrons scattering out of the active region of the MUSIC chamber. A typical cut containing 89\% of all $^{198}$Pb$^{+80}$ in the beam had a purity of 86\%. The maximum purity of a cut is limited by any charge states of the same element that may overlap in velocity with the isotope of interest. If there is no \emph{directly} overlapping charge state then a cut with an arbitrarily high purity can be made at the cost of lost statistics. In the case of $^{198}$Pb$^{+80}$, a cut containing less than 20\% of all $^{198}$Pb$^{+80}$ in the beam with purity greater than 99\% can be achieved. 

Using isomer tagging, the charge state distributions of the isotopes in the 74 MeV/A radioactive ion beams were measured. These distributions were compared to calculations done with the GLOBAL model using LISE++ to simulate the beam optics. GLOBAL was found to under-predict the electron capture cross-sections; nevertheless, good agreement between experiment and model was achieved by adjusting the magnetic rigidity (and therefore velocity) of the beam fragments in the simulation.

\begin{acknowledgments}
We would like to acknowledge support from Michigan State University, from the U.S Department of Energy under Grant Nos. DE-SC0020451, DE-SC0014552, and DE-NA0003908, from the National Science Foundation under Grant Nos. PHY-1565546, and PHY-1712832.
\end{acknowledgments}

\section*{Data Availability Statement}
The data that support the findings of this study are available from the corresponding author upon reasonable request.

%
%

%


\bibliography{main.bib}

\end{document}